\documentclass[12pt,pre,aps,,showpacs,preprint]{revtex4}

\usepackage{psfig}
\usepackage{amsfonts}
\begin{document}

\title{Exactly Solvable Disordered Sphere-Packing Model
           in Arbitrary-Dimension Euclidean Spaces}

\author{S. Torquato}

\email{torquato@electron.princeton.edu}

\affiliation{\emph{Department of Chemistry}, \emph{Princeton University}, Princeton
NJ 08544}

\affiliation{\emph{Program in Applied and Computational Mathematics}, \emph{Princeton
University}, Princeton NJ 08544}

\affiliation{\emph{PRISM, Princeton University}, Princeton NJ 08544}

\author{F. H. Stillinger}

\affiliation{\emph{Department of Chemistry}, \emph{Princeton University}, Princeton
NJ 08544}

\begin{abstract}

We introduce a generalization of the well-known random sequential addition (RSA) process
for hard spheres in $d$-dimensional Euclidean space $\mathbb{R}^d$. We show that
all of the $n$-particle correlation functions 
of this nonequilibrium model, in a certain limit called the ``ghost" RSA packing, can be obtained analytically for 
all allowable densities and in any dimension. This represents the first exactly
solvable disordered sphere-packing model in arbitrary dimension. 
The fact that the maximal density $\phi(\infty)=1/2^d$ of the ghost RSA  packing implies
that there may be disordered sphere packings in sufficiently high $d$ whose density exceeds
Minkowski's lower bound  for Bravais lattices, the dominant asymptotic
term of which is $1/2^d$. Indeed, we report on
a conjectural lower bound on the density whose asymptotic behavior
is controlled by $2^{-(0.77865\ldots) d}$, thus providing the putative
exponential improvement on Minkowski's 100-year-old bound. 
Our results suggest that the densest packings in sufficiently
high dimensions may be disordered rather than periodic, implying
the existence of disordered classical ground states for some continuous potentials.

\end{abstract}
\pacs{05.20.-y, 61.20.-p}

\maketitle

\section{Introduction}

A collection of congruent spheres in $d$-dimensional Euclidean space $\mathbb{R}^d$ 
is called a sphere packing if no two spheres overlap.
The {\it packing density} or simply density $\phi$ of a sphere packing is the fraction of
space $\mathbb{R}^d$ covered by the spheres.
Hard-sphere packings have been used to model a variety of systems, including liquids \cite{Ha86},
amorphous and granular media \cite{To02a}, and crystals \cite{Chaik95}. Nonetheless, 
there is great interest in understanding sphere packings
in high dimensions in various fields. For example, it is known 
that the optimal way of sending digital signals over noisy channels correspond 
to the densest sphere packing in a high dimensional space \cite{Co93}.
These ``error-correcting" codes underlie a variety of systems in digital
communications and storage, including compact disks, cell phones and the Internet.
Physicists have studied hard-sphere packings in high dimensions to gain insight
into ground and glassy states of matter as well as 
phase behavior in lower dimensions \cite{Fr99,Pa00}. The determination of the densest packings
in arbitrary dimension is a problem of long-standing interest in discrete geometry \cite{Co93}.

It is instructive to note that upper and
lower bounds on the {\it maximal density}
\begin{equation}
\phi_{\mbox{\scriptsize max}}= \sup_{P \subset \Re^d} \phi(P)
\end{equation}
exist in all dimensions \cite{Co93}, where the supremum is taken over
all packings $P$ in $\mathbb{R}^d$.
For example, Minkowski \cite{Mi05} proved that the maximal density 
$\phi^L_{\mbox{\scriptsize max}}$ among all Bravais lattice packings 
for $d \ge 2$ satisfies the lower bound
\begin{equation}
\phi^L_{\mbox{\scriptsize max}} \ge \frac{\zeta(d)}{2^{d-1}},
\label{mink}
\end{equation}
where $\zeta(d)=\sum_{k=1}^\infty k^{-d}$ is the Riemann zeta function.
One observes that for large values of $d$,
the asymptotic behavior of the {\it nonconstructive} Minkowski lower bound is controlled by $2^{-d}$.
Interestingly,
the density of a {\it saturated} packing of congruent spheres
in $\mathbb{R}^d$ for all $d$ satisfies 
\begin{equation}
\phi \ge \frac{1}{2^d}.
\label{sat}
\end{equation}
A saturated packing of congruent spheres
of unit diameter  and density $\phi$ in $\Re^d$ has the property that each point in space lies
within a unit distance from the center of some sphere. Thus, a covering
of the space is achieved if each  center is encompassed by a sphere
of unit radius  and the density of this covering is $2^d \phi \ge 1$, which proves the 
so-called {\it greedy} lower bound (\ref{sat}). Note that it has  the same 
dominant exponential term as (\ref{mink}).

A statistically homogeneous (i.e., translationally invariant) packing is completely configurationally characterized
by specifying all of the $n$-particle correlation functions.
For such packings in $\mathbb{R}^d$, these correlation functions 
are defined so that $\rho^n g_n({\bf r}_1,{\bf r}_2,\dots,{\bf r}_n)$  is proportional to
the probability density for simultaneously finding $n$ particles at 
locations ${\bf r}_1,{\bf r}_2,\dots,{\bf r}_n$ within the system,
where $\rho$ is the number density. Thus, each $g_n$ approaches
unity when all particle positions become widely separated within $\mathbb{R}^d$,
indicating no spatial correlations.
To date, an exact determination of all of the $n$-particle correlation functions for a 
packing has only been possible for $d=1$ in the special case of an equilibrium ensemble
of such particles \cite{Sa53}. Observe that in the limit $d \rightarrow \infty$,
it is known that the pressure of an equilibrium hard-sphere fluid 
is exactly given by the low-density expansion up to the second-virial level for
a positive range of densities \cite{Fr99},
which implies a simplified form for all of the correlation functions \cite{To05b}.

We present in Section II a generalization of the well-known random sequential
addition (RSA) process of hard particles \cite{Re63,To02a}. In a particular limit 
of this nonequilibrium model that we call the ``ghost" RSA process, we are able to obtain the $g_n$ for
all allowable densities exactly for any $n$ and dimension $d$.
The key geometric quantity that determines $g_n$ is the union volume of $n$ overlapping
exclusion spheres of radius equal to the sphere diameter. 
We show that this construction of
a disordered but {\it unsaturated} packing realizes the greedy 
lower bound (\ref{sat}). This implies
that there may be disordered sphere packings in sufficiently high $d$ whose density exceeds
Minkowski's lower bound (\ref{mink}). Indeed, in Section III, we report on
a conjectural lower bound on the density whose asymptotic behavior
is controlled by $2^{-(0.77865\ldots) d}$, thus providing the putative
exponential improvement on Minkowski's 100-year-old bound. 
Our results lead to the counterintuitive possibility that
optimal packings in sufficiently high dimensions may be disordered
and thus have implications
for our fundamental understanding of classical ground states of matter.

\section{Generalized Random Sequential Addition Model}

We introduce a disordered sphere-packing model in $\mathbb{R}^d$ that is a subset of
the Poisson point process and is a generalization of the standard random RSA process. 
The centers of ``test" spheres of unit diameter arrive continually 
throughout $\mathbb{R}^d$ during time $t\ge 0$ according to a translationally invariant Poisson process
of density $\eta$ per unit time, i.e., $\eta$ is the number of
points per unit volume and time. Therefore, the expected number of
centers in a region of volume $\Omega$ during time $t$ is $\eta \Omega t$
and the probability that this region is empty of centers is $\exp(-\eta \Omega t)$.
However, this Poisson distribution of test spheres is not a packing because 
the spheres can overlap. To create a packing from this point process,
one must remove test spheres such that no sphere center can lie within a spherical 
region of unit radius from any sphere center. Without loss of generality,
we will set $\eta=1$.

There is a variety of ways of achieving this ``thinning" process such that the subset of points
correspond to a sphere packing. One obvious rule is to retain a test
sphere at time $t$ only if it does not overlap a sphere that was successfully
added to the packing at an earlier time. This criterion defines the standard RSA process in $\mathbb{R}^d$
\cite{To02a,Re63}, which generates a homogeneous and isotropic
sphere packing in $\mathbb{R}^d$ with a time-dependent density
$\phi(t)$. In the limit $t \rightarrow \infty$, the RSA process corresponds to a saturated
packing with a maximal or {\it saturation} density $\phi_s(\infty) \equiv \lim_{t\rightarrow
\infty} \phi(t)$. In one dimension, the RSA process is commonly known as the ``car parking problem", 
which Re{\' n}yi showed has a saturation density $\phi_s(\infty)= 0.7476\ldots$ \cite{Re63}.
For $2 \le d < \infty$, an exact determination of $\phi_s(\infty)$ is not
possible, but estimates for it have been obtained via computer experiments
for low dimensions \cite{To02a}. 

\begin{figure}
\centerline{\psfig{file=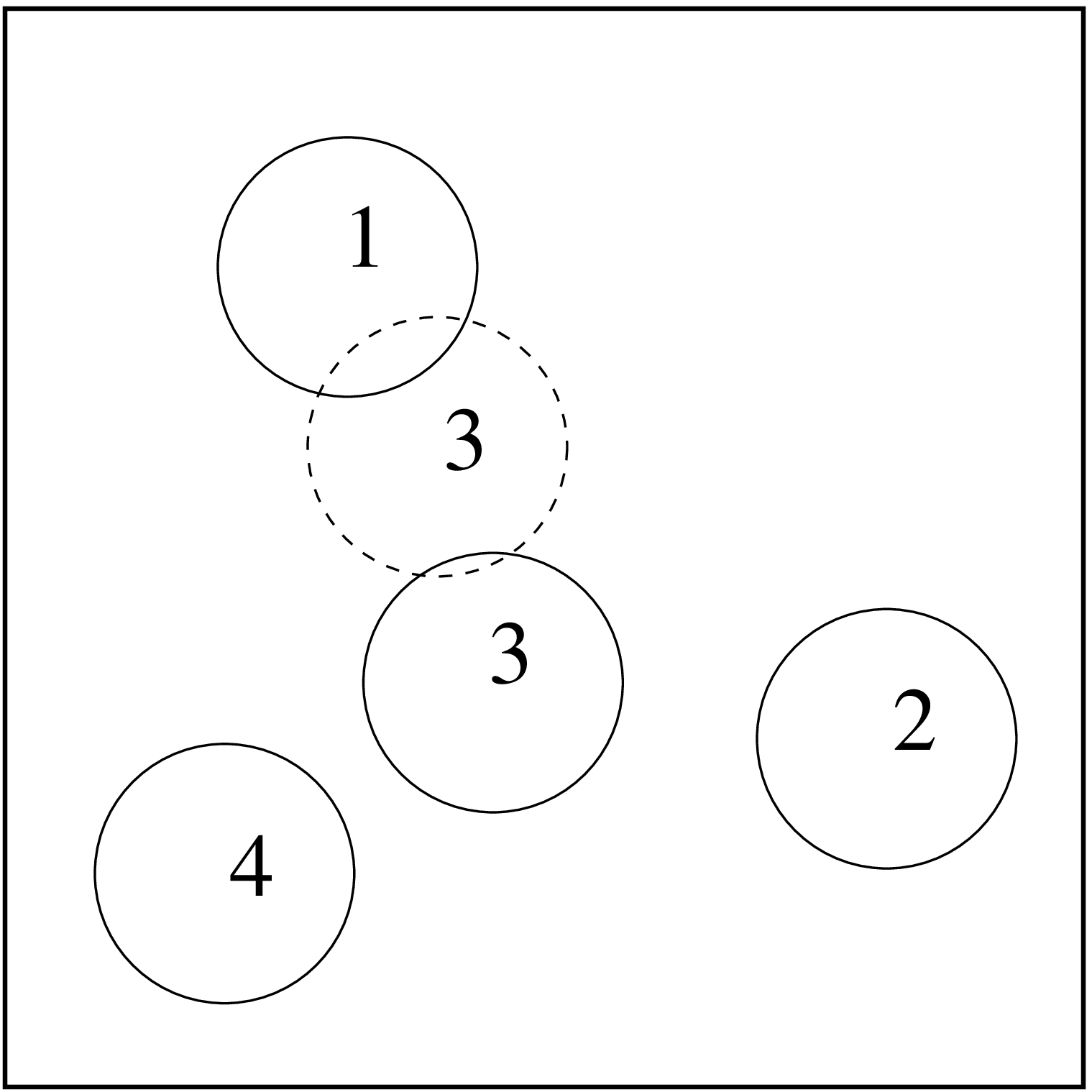,height=1.5in}
\hspace{0.2in}\psfig{file=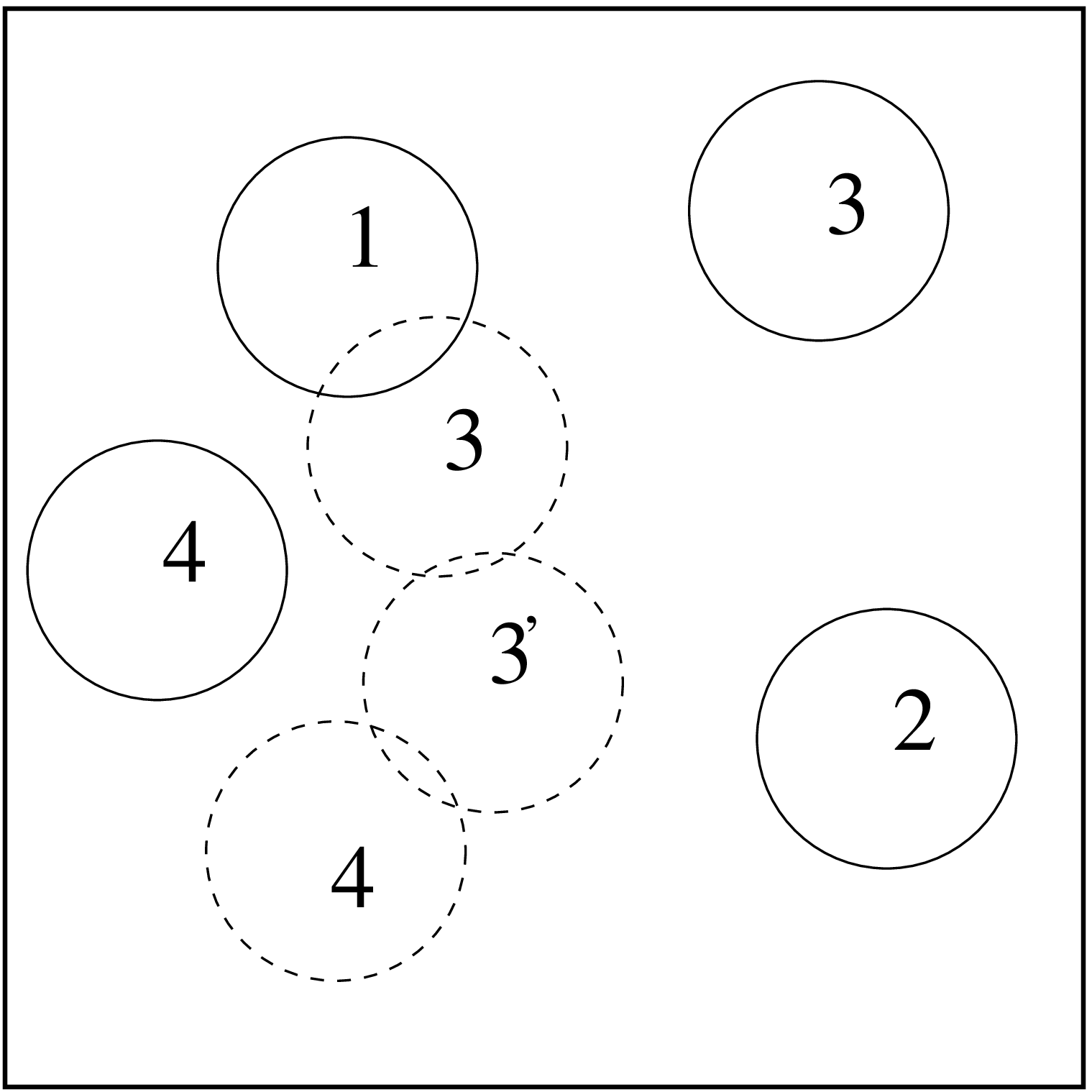,height=1.5in} }
\caption{The addition of four successfully added particles (in the numerical
order indicated) in the generalized RSA process at the
two extremes of $\kappa=0$ (left panel) and $\kappa=1$ (right panel).
In both cases, the rejected particles have dashed boundaries. For the
case $\kappa=1$, a test sphere cannot overlap a ghost sphere. 
Here $3^\prime$ represents the second attempt to add a third sphere.}
\label{processes}
\end{figure}

Another thinning criterion retains a test sphere centered at position $\bf r$ at time $t$ 
if no other test sphere is within a unit radial distance from $\bf r$ for
the time interval $\kappa t$ prior to $t$, where $\kappa$ is a positive constant
in the interval $[0,1]$. This packing is a subset of the RSA packing, and
hence we refer to it as the generalized RSA process. Note
that when $\kappa=0$, the standard RSA process is recovered, and when $\kappa=1$,
a model due to Mat{\' e}rn \cite{Ma86} is recovered \cite{footnote1}. The latter is amenable
to exact analysis and is the main focus of this paper. 
For any $0 < \kappa \le 1$, the generalized
RSA process is always an {\it unsaturated} packing.
Figure \ref{processes} illustrates the differences
between the generalized RSA process at the
two extremes of $\kappa=0$ and $\kappa=1$.
In remainder of this section, we will focus on the case $\kappa=1$.

The time-dependent density $\phi(t)$ in the case of the generalized RSA process
with $\kappa=1$ is easily obtained. In this packing,
a test sphere at time $t$ is retained only if does not overlap 
an existing sphere in the packing as well as any previously rejected
test sphere, which we will call ``ghost" spheres. The model itself
will be referred to as the ghost RSA process. An overlap cannot
occur if a test sphere is outside a unit radius of any successfully
added sphere or ghost sphere. Because of the underlying Poisson process,
the probability that a trial sphere is retained 
at time $t$ is given by $\exp(-v_1(1) t)$, where $v_1(1)=\pi^{d/2}/\Gamma(1+d/2)$ is the volume
of a sphere of unit radius. Therefore, the expected time-dependent number density $\rho(t)$ and packing density 
$\phi(t)=\rho(t)v_1({1/2})$ at any time $t$ are given by
\begin{equation}
\rho(t)=\int_0^t \exp(-v_1(1) t^\prime) dt^\prime=\frac{1-\exp(-v_1(1) t)}{v_1(1)},
\quad \phi(t)=\frac{1-\exp(-v_1(1) t)}{2^d}.
\label{rho(t)}
\end{equation}
In the limit $t \rightarrow \infty$, we therefore have that
\begin{equation}
\rho(\infty) \equiv \lim_{t \rightarrow \infty} \rho(t)=\frac{1}{v_1(1)}, \quad \phi(\infty) \equiv \lim_{t \rightarrow \infty} \phi(t)=\frac{1}{2^d}.
\label{rho}
\end{equation}
Observe that the greedy lower bound (\ref{sat}) on the density is achieved in 
the infinite-time limit for this sequential but {\it unsaturated} packing,
which was pointed out only recently \cite{To05b}.
Although the limiting packing density $\phi(\infty)=1/2^d$ is far
from optimal in low dimensions, it is relatively large in high dimensions,
as discussed in our concluding remarks. Obviously, for any $0 \le \kappa < 1$, the maximum (infinite-time)
 density of the generalized RSA packing is bounded from below by $1/2^d$
(i.e., the maximum density for $\kappa=1$).  Henceforth, we write $v_1\equiv v_1(1)$.

\begin{figure}[bthp]
\centerline{\psfig{file=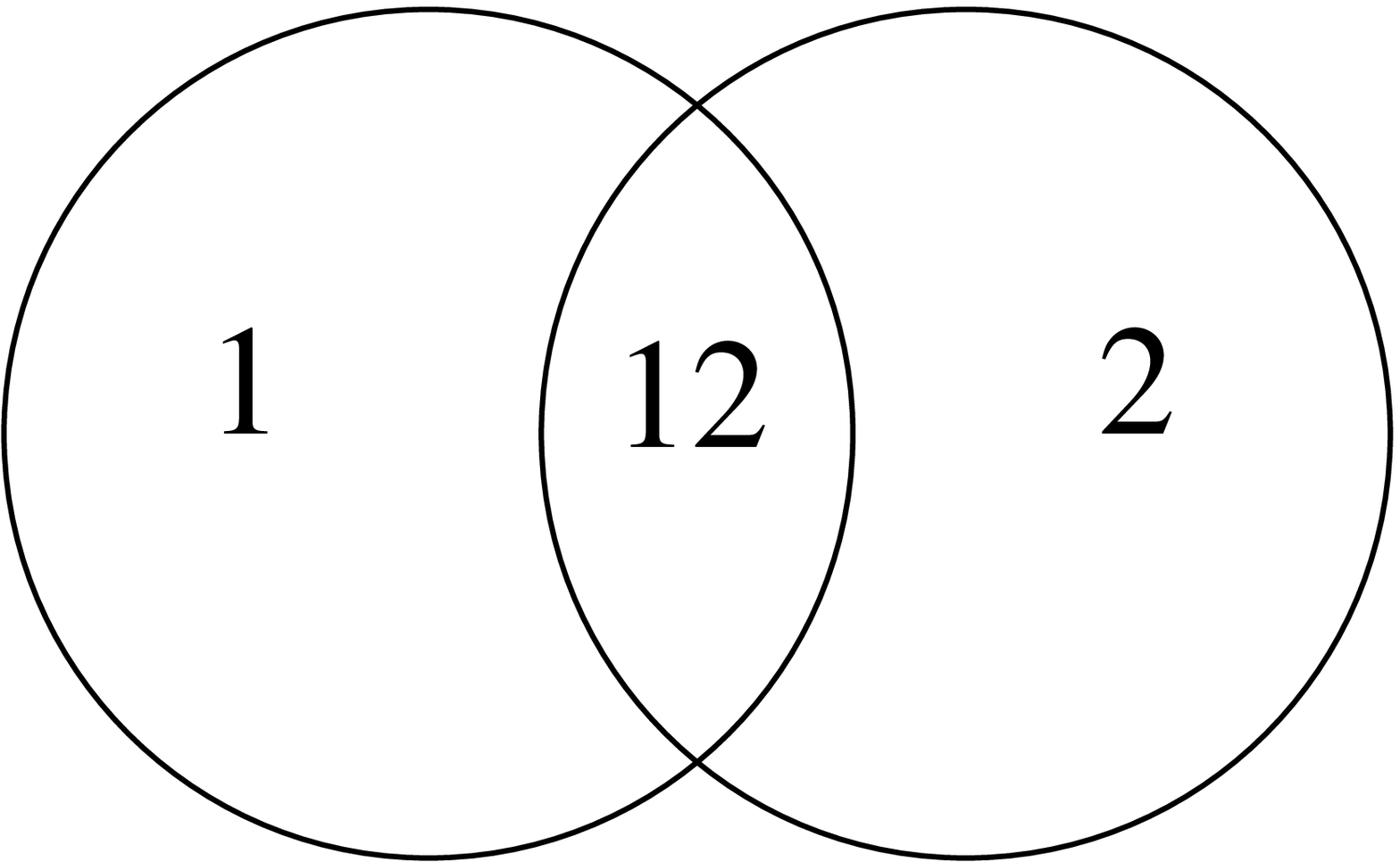,height=0.9in}
\hspace{0.75in}\psfig{file=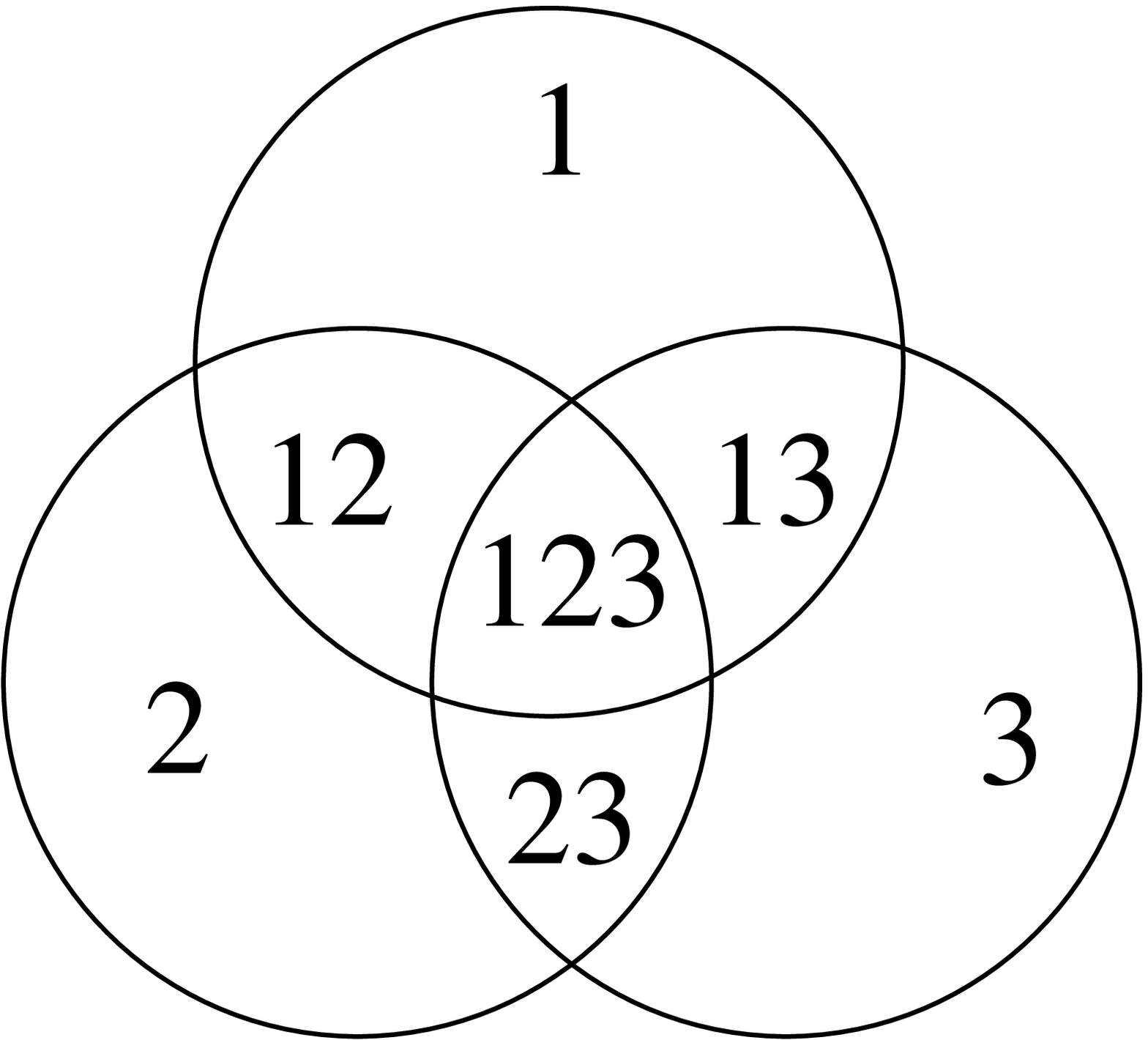,height=0.9in}}
\caption{ Left panel: Relevant subvolumes for two overlapping spheres of unit radius
associated with the arrivals of two test spheres. The labels refer to
distinct, nonoverlapping regions.
Right panel: Relevant subvolumes for three overlapping spheres of unit radius
associated with the arrivals of three test spheres.  }
\label{2-int}
\end{figure}

The derivation of the expression of $g_2(r;t)$ is actually a simple extension of the 
aforementioned one for $\rho(t)$. Two test spheres that arrive at times $t_1$ and $t_2$ 
and whose centers are separated by a distance $r$ can only be retained if no other test 
spheres arrived before $t_1$ and $t_2$, respectively (see Fig. \ref{2-int}). 
Thus, the key geometrical object is 
the union volume $v_2(r)$ of two spheres of unit radius whose centers are separated by a 
distance $r$, which can be expressed in terms of the intersection volume $v_2^{int}(r)$ 
\cite{footnote2} between two such spheres via the relation 
\begin{displaymath}
v_2(r)=2 v_1 -v_2^{int}(r).
\end{displaymath}
For $r \ge 2$, there is no volume common to two such spheres ($v_2^{int}(r)=0$) and therefore 
$g_2(r;t)=1$, i.e., pair correlations vanish.
However, if $1 \le r \le 2$, the two spheres have a common volume and 
\begin{eqnarray}
\rho^2(t)g_2(r;t)&=&\int_0^t \int_0^t \exp\Big[-t_1[v_1-v_2^{int}(r)] 
-t_2[v_1-v_2^{int}(r)]  -\max{(t_1,t_2)} v_2^{int}(r) \Big]dt_1dt_2 \nonumber \\
&=& 2\int_0^t dt_2\exp[-t_2 v_1] \int_0^{t_2} dt_1 \exp\Big[-t_1[v_1-v_2^{int}(r)]\Big]  \nonumber \\
&=& \frac{2}{v_2(r)-v_1}\Big[\frac{1-e^{-tv_1}}{v_1}-\frac{1-e^{-tv_2(r)}}{v_2(r)}\Big].
\label{g2-1}
\end{eqnarray}
In relation (\ref{g2-1}), the terms within the first three brackets are the 
distinct volumes of the regions
labeled 1, 2 and 12 in the left panel of Fig. \ref{2-int}.
Therefore, the time-dependent
pair correlation function for all $r$ and $t$ is given by
\begin{equation}
\rho^2(t)g_2(r;t)= \frac{2\Theta(r-1)}{v_2(r)-v_1}
\Big[ \frac{1-e^{-tv_1}}{v_1}-\frac{1-e^{-tv_2(r)}}{v_2(r)}\Big],
\label{g2-2}
\end{equation}
where $\Theta(x)$ is the unit step function, 
equal to zero for $x<0$ and unity for $x \ge1$.
It is useful to note that at small times or, equivalently, low densities,
formula (\ref{rho(t)}) yields the asymptotic expansion
$\phi(t)=t- 2^{d-1} t^2+{\cal O}(t^3)$,
which when inverted yields $t=\phi+ 2^{d-1} \phi^2+{\cal O}(\phi^3)$.
Substitution of this last result into (\ref{g2-2}) gives
\begin{equation}
g_2(r;\phi)=\Theta(r-1)+ {\cal O}(\phi^3),
\label{g2-t-grsa}
\end{equation}
which implies that $g_2(r;\phi)$ tends to the unit step function
$\Theta(r-1)$ as $\phi \rightarrow 0$ for any $d$.

\begin{figure}[bthp]
\centerline{\psfig{file=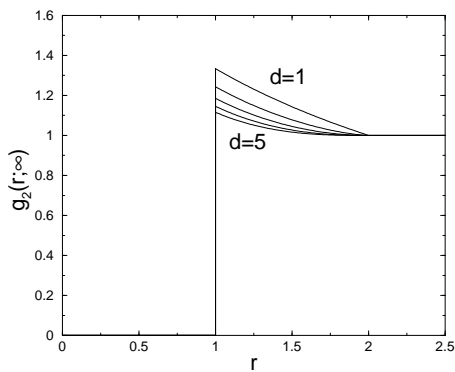,height=2.0in}}
\caption{ Radial distribution function for the first five space dimensions
at the maximum density $\phi=1/2^d$ for the generalized RSA model with $\kappa=1$, i.e.,
the ``ghost RSA process."}
\label{grsa}
\end{figure}

In the limit $t \rightarrow \infty$, we have from (\ref{g2-2}) that
$\rho^2(\infty)g_2(r;\infty)=2\Theta(r-1)/[v_1v_2(r)]$
or, using (3), 
\begin{equation}
g_2(r;\infty)=\frac{2\Theta(r-1)}{\beta_2(r)},
\label{g2-3}
\end{equation}
where $\beta_2(r)=v_2(r)/v_1$. 
The radial distribution function $g_2(r;\infty)$ is plotted in Fig. \ref{grsa}
for the first five space dimensions.
Because $\beta_2(r)$ is equal to 2 for $r \ge 2$,  
$g_2(r;\infty)=1$ for $r \ge 2$, i.e., spatial correlations
vanish identically for all pair distances except
those in the small interval $[0,2)$. Even the positive correlations
exhibited for $1 < r <2$ are rather weak and decrease 
exponentially fast with increasing dimension \cite{To05b}, i.e.,  $g_2(r;\infty)$ tends  to the unit step 
function as $d \rightarrow \infty$, i.e., beyond the hard core (a constrained
correlation), spatial correlations
vanish. 

Mat{\' e}rn  originally gave an expression for the time-dependent density $\phi(t)$ and
and a formal expression (as opposed to explicit expression for any $d$) for
the time-dependent radial distribution function $g_2(r;t)$ when $\kappa=1$
using a completely different approach. However, he did not consider obtaining any
of the higher-order correlation functions.

Let us now derive the time-dependent triplet correlation function
 $g_3({\bf r}_{12},{\bf r}_{13};\infty)$.
Here the relevant geometrical object
is the union volume $v_3(r_{12},r_{13},r_{23})$ of three spheres of unit radius whose centers are
separated by the distances $r_{12}$, $r_{13}$ and $r_{23}$, which can be expressed in terms 
of the intersection volume $v_3^{int}(r_{12},r_{13},r_{23})$ between three such spheres via
the relation
\begin{equation}
v_3(r_{12},r_{13},r_{23})=3 v_1 -v_2^{int}(r_{12}) -v_2^{int}(r_{13}) 
-v_2^{int}(r_{23}) +v_3^{int}(r_{12},r_{13},r_{23}).
\end{equation}
Whenever there is no overlap between the three spheres, $g_3=1$, i.e., triplet correlations vanish. On the
other hand, whenever the spheres overlap such that each pair distance is
greater than or equal to unity, there are triplet correlations.
In such situations, it is convenient to introduce
the time-dependent triplet function
\begin{eqnarray}
&&F(r_{12},r_{13},r_{23};t_1,t_2,t_3)= -t_1[v_1-v_2^{int}(r_{12})-v_2^{int}(r_{13})
+ v_3^{int}(r_{12},r_{13},r_{23})] \nonumber \\
&& -t_2[v_1-v_2^{int}(r_{12})-v_2^{int}(r_{23})+v_3^{int}(r_{12},r_{13},r_{23})] 
-t_3[v_1-v_2^{int}(r_{13})-v_2^{int}(r_{23})+v_3^{int}(r_{12},r_{13},r_{23})] \nonumber \\ 
&&-\max{(t_1,t_2)}[v_2^{int}(r_{12})-v_3^{int}(r_{12},r_{13},r_{23})] 
-\max{(t_1,t_3)}[v_2^{int}(r_{13})-v_3^{int}(r_{12},r_{13},r_{23})] \nonumber \\
&&-\max{(t_2,t_3)}[v_2^{int}(r_{23})-v_3^{int}(r_{12},r_{13},r_{23})] 
-\max{(t_1,t_2,t_3)}v_3^{int}(r_{12},r_{13},r_{23}).
\end{eqnarray}
The terms within the first three brackets are the volumes of the regions
labeled 1, 2 and 3 in the right panel of Fig. \ref{2-int}. The terms within the fourth through
sixth  brackets are the volumes labeled 12, 13 and 23 in the right panel of Fig. \ref{2-int}.
Of course, the region labeled 123 denotes the intersection volume of three spheres. 
The triplet correlation function at time $t$ is
given by
\begin{displaymath}
\rho^3(t)g_3(r_{12},r_{13},r_{23};t)=\int_0^t \int_0^t \int_0^t
 \exp\Big[-F(r_{12},r_{13},r_{23};t_1,t_2,t_3)   \Big]dt_1dt_2 dt_3
\end{displaymath}
and therefore at infinitely large times we have, using  (\ref{rho}, (\ref{g2-3}) and (\ref{g3-1}), that
\begin{eqnarray}
&&\rho^3(\infty)g_3(r_{12},r_{13},r_{23};\infty)= 2\int_0^\infty dt_3\exp[-t_3 v_1] \int_0^{t_3} dt_2 
\exp\Big[-t_2[v_1-v_2^{int}(r_{23})]\Big] \nonumber \\ &&\times 
 \int_0^{t_2} dt_1 \exp\Big[-t_1[v_1-v_2^{int}(r_{12})-v_2^{int}(r_{13})+v_3^{int}(r_{12},r_{13},r_{23})] \Big] \nonumber \\
&+& 2\int_0^\infty dt_1\exp[-t_1 v_1] \int_0^{t_1} dt_3 \exp\Big[-t_3[v_1-v_2^{int}(r_{13})]\Big] \nonumber \\
 &&\times 
 \int_0^{t_3} dt_2 \exp\Big[-t_2[v_1-v_2^{int}(r_{12})-v_2^{int}(r_{23})+v_3^{int}(r_{12},r_{13},r_{23})] \Big] \nonumber \\
 &+& 2\int_0^\infty dt_2\exp[-t_2 v_1] \int_0^{t_2} dt_1 \exp\Big[-t_1[v_1-v_2^{int}(r_{12})]\Big] \nonumber \\
 &&\times 
 \int_0^{t_1} dt_3
 \exp\Big[-t_3[v_1-v_2^{int}(r_{13})-v_2^{int}(r_{23})+v_3^{int}(r_{12},r_{13},r_{23})] \Big]
 \nonumber\\
 &=& \frac{2}{v_1v_3(r_{12},r_{13},r_{23})}\Bigg[\frac{1}{v_2(r_{12})}+
 \frac{1}{v_2(r_{13})}+\frac{1}{v_2(r_{23})}\Bigg].
\end{eqnarray}
Combination of (\ref{rho}, (\ref{g2-3}) and (\ref{g3-1}) yields
the following expression for the triplet correlation function for arbitrary
positions at infinitely large times:
\begin{eqnarray} 
g_3({\bf r}_{12},{\bf r}_{13};\infty)&=&
\frac{\Theta(r_{12}-1)\Theta(r_{13}-1)\Theta(r_{23}-1)}{\beta_3(r_{12},r_{13},r_{23})}
\Big[g_2(r_{12};\infty)+g_2(r_{13};\infty)+g_2(r_{23};\infty)\Big],
\label{g3-1}
\end{eqnarray}
where $\beta_3(r_{12},r_{13},r_{23})=v_3(r_{12},r_{13},r_{23})/v_1$ and 
$g_3({\bf r}_{12},{\bf r}_{13};\infty) \equiv g_3(r_{12},r_{13},r_{23};\infty)$.

A similar analysis reveals that the four-particle correlation function
in the limit $t \rightarrow \infty$ is given by
\begin{equation}
g_4({\bf r}_{12},{\bf r}_{13},{\bf r}_{14};\infty)=\frac{\prod_{i<j}^4
\Theta(r_{ij}-1)}{\beta_4({\bf r}_{12},{\bf r}_{13},{\bf r}_{14};\infty)}\Big[
g_3({\bf r}_{12},{\bf r}_{13};\infty)+g_3({\bf r}_{12},{\bf r}_{14};\infty)+
g_3({\bf r}_{13},{\bf r}_{14};\infty)+g_3({\bf r}_{23},{\bf r}_{24};\infty)\Big]
\end{equation}
By induction, the $n$-particle  correlation function for arbitrary
positions at infinitely large times is given by
\begin{equation}
g_n({\bf r}_{1}, \ldots, {\bf r}_{n};\infty)= 
\frac{\prod_{i<j}^n\Theta(r_{ij}-1)}{
\beta_n({\bf r}_{1},\ldots,{\bf r}_{n})}
\Big[\sum_{i=1}^{n}g_{n-1}(Q_i;\infty)\Big],
\label{gn-grsa}
\end{equation}
where the sum is over  all the $n$ distinguishable ways of choosing
$n-1$ positions from $n$ positions ${\bf r}_1,\ldots {\bf r}_n$ and the arguments of $g_{n-1}$
are the associated $n-1$ positions, which we denote by $Q_i$.
 Moreover, $\beta_n({\bf r}_{12},{\bf r}_{13},\ldots,{\bf r}_{1n})=
v_n({\bf r}_{12},{\bf r}_{13},\ldots,{\bf r}_{1n})/v_1$,
where $v_n({\bf r}_{12},{\bf r}_{13},\ldots,{\bf r}_{1n})$ is the union volume
of $n$ congruent spheres of unit radius whose centers are located at
${\bf r}_1,\ldots, {\bf r_n}$, ${\bf r}_{ij}={\bf r}_j -{\bf r}_i$
for all $1 \le i <j \le n$
and $r_{ij} =|{\bf r}_{ij}|$.

It can be shown \cite{To05b}
that in the limit $d \rightarrow \infty$ and for $\phi=1/2^d$
\begin{equation}
g_n({\bf r}_{12}, \ldots, {\bf r}_{1n};\infty) \sim \prod_{i<j}^n g_2(r_{ij};\infty),
\end{equation}
where $g_2(r;\infty) \sim \Theta(r-1)$.
We see that unconstrained spatial correlations vanish asymptotically.
Specifically, (i) the high-dimensional asymptotic behavior of $g_2$ is the same as the asymptotic
behavior in the low-density limit for any $d$ [cf. (\ref{g2-t-grsa})], 
i.e., {\it unconstrained} spatial
correlations, which exist for positive densities at fixed $d$, vanish
asymptotically for pair distances beyond the hard-core diameter in the high-dimensional limit; 
and (ii) $g_n$ for $n \ge 3$ asymptotically can be inferred from
a knowledge of only the pair correlation function $g_2$ and number density $\rho$.
These two asymptotic properties, 
which we have called the {\it decorrelation principle} \cite{To05b},
apply more generally to any disordered packing, as discussed in Ref. \cite{To05b}.
Asymptotically, unconstrained correlations vanish (i.e., statistical
independence is established) because we know from  
the Kabatiansky and Levenshtein  asymptotic upper bound on the maximal density
$\phi_{\mbox{\scriptsize max}}$ of any sphere packing that the density must go to zero at least 
as fast as $2^{-0.5990d}$ for large $d$ \cite{Ka78}.

\section{Discussion}

The fact that the maximal density $\phi(\infty)=1/2^d$ of the ghost RSA  packing 
coincides with the greedy lower bound (\ref{sat}) strongly suggests that there
are saturated disordered packings that have larger densities, i.e., the greedy
lower bound is a weak bound for saturated packings \cite{footnote3}. This implies
that there may be disordered sphere packings in sufficiently high $d$ whose density exceeds
Minkowski's lower bound (\ref{mink}) for Bravais lattices, the dominant asymptotic
term of which is $1/2^d$. Our results
already give insight into this fascinating possibility. For example, consider the so-called
checkerboard lattice $D_d$ in $d$ dimensions \cite{Co93}, which is a $d$-dimensional
generalization of the optimal (densest) face-centered cubic lattice in three dimensions,
and thought to be the optimal packing for $d=4$ and $d=5$. Its packing density
$\phi=\pi^{d/2}/[\Gamma(1+d/2)2^{(d+2)/2}]$ exponentially decreases with 
increasing $d$ (because it quickly becomes
unsaturated) and falls below the ghost-RSA-process value of $1/2^d$ for the first time
at $d=28$ \cite{footnote4}. The ratio of densities of the ghost RSA process to the
checkerboard  at $d=100$ is given by $\phi_{ghost}/\phi_{checker} \approx 7.5\times
10^{25}$. Although both packings are unsaturated in such high dimensions,
the fact that $g_2(r)$ for the ghost RSA process is effectively uniform (unity) 
for all $r>1$  but for the checkerboard lattice involves Dirac delta functions
of {\it weak} strength at {\it widely spaced} discrete distances explains why the former 
is enormously denser than the latter. 

Over the last century, many extensions and generalizations of Minkowski's
lower bound (\ref{mink}) have been obtained \cite{Co93}, but none of these 
investigations have been able to improve upon the dominant exponential term $2^{-d}$. In another
work \cite{To05b}, we will present comprehensive rigorous evidence that this exponential improvement
may be provided by considering specific disordered sphere packings.
Here we simply sketch the procedure leading to this putative improvement over
Minkowski's lower bound. The basic ideas underlying our new approach to the derivation of lower bounds on 
$\phi_{\mbox{\scriptsize max}}$ were actually described in our earlier work \cite{To02c}
in which we studied so-called $g_2$-invariant processes.
A {\it $g_2$-invariant process} is one in which a given nonnegative pair correlation 
$g_2({\bf r})$ function remains invariant for all ${\bf r}$ over the range of
densities
\begin{equation}
0 \le \phi \le \phi_*.
\end{equation}
The terminal density $\phi_*$ is the maximum achievable density
for the $g_2$-invariant process subject to satisfaction of
certain necessary conditions on the pair correlation. 
In particular, we considered those ``test" $g_2(r)$'s that are distributions on $\mathbb{R}^d$ depending
only on the radial distance $r$.
For any test $g_2(r)$, we want to maximize 
the corresponding density $\phi$ satisfying the following three conditions:

\noindent (i) \hspace{0.25in} $g_2(r)  \ge 0 \qquad \mbox{for all}\quad  r,$

\noindent (i) \hspace{0.2in} $g_2(r)= 0 \qquad \mbox{for}\quad  r<1,$

\noindent (iii)
\begin{displaymath}
\hspace{-0.5in}S(k)= 1+ \rho\left(2\pi\right)^{\frac{d}{2}}\int_{0}^{\infty}r^{d-1}h(r)
\frac{J_{\left(d/2\right)-1}\!\left(kr\right)}{\left(kr\right)^{\left(d/2\right
)-1}}dr \ge 0  \qquad \mbox{for all}\quad  k,
\end{displaymath}
where $h(r)=g_2(r)-1$ is the total correlation function.
Condition (i) is a trivial consequence of the fact that $g_2$ is a probability density
function. Condition (ii) is just the {\it hard-core} constraint for
spheres of unit diameter. Condition (iii)
states that the structure factor $S(k)$ in $d$ dimensions 
must be nonnegative for all $k$.
When there exist sphere packings with $g_2$ satisfying conditions
(i)-(iii) for $\phi$ in the interval $[0,\phi_*]$, then we have the lower
bound on the maximal density given by
\begin{equation}
\phi_{\mbox{\scriptsize max}} \ge \phi_*.
\label{true-bound-phi}
\end{equation}

It is rather remarkable that the optimization problem defined above is identical
to one formulated by Cohn \cite{Co02}. Specifically, it is the {\it dual}
of the {\it primal} infinite-dimensional linear program that Cohn employed with Elkies \cite{Co03}
to obtain upper bounds on the maximal packing density. Thus, even if  there does 
not exist a sphere packing with $g_2$ satisfying conditions
(i)-(iii), the terminal density $\phi_*$ can never exceed the
Cohn-Elkies upper bound and, more generally, our formulation has implications for upper bounds
on $\phi_{\mbox{\scriptsize max}}$.

In addition, to the 
structure factor condition, there are generally many other conditions that a pair correlation
function corresponding to a point process must obey \cite{Cos04}.
One such additional necessary condition, obtained by Yamada \cite{Ya61},
is concerned with the variance $\sigma^2(\Omega) \equiv 
\langle (N(\Omega)^2- \langle N(\Omega) \rangle)^2\rangle$,
in the number $N(\Omega)$ of particle centers  contained within a region or ``window"
$\Omega \subset \mathbb{R}^d$:
\begin{equation}
\sigma^2(\Omega)=\rho |\Omega| \Big[ 1+\rho \int_{\Omega} h({\bf r}) d{\mathbf r}\Big] \ge \theta(1-\theta),
\label{yamada}
\end{equation}
where $\theta$ is the fractional part of the expected number of points 
$\rho |\Omega|$ contained in
the window. This is a consequence of the fact that the number of particles
in any window must be an integer.

In Ref. \cite{To02c}, a five-parameter test family of $g_2$'s had been considered,
which incorporated the known features of core exclusion, contact pairs, and damped oscillatory short-range
order beyond contact that are features intended to 
describe {\it disordered} jammed sphere packings for $d=3$.   However, because
of the functional complexity of this test $g_2$, the terminal density could only be determined numerically.
The general optimization procedure outlined above was employed in Ref. \cite{To05b} to obtain
analytical estimates of the terminal density in high dimensions that together with 
the following conjecture provide the putative exponential improvement
on Minkowski's lower bound on $\phi_{\mbox{\scriptsize max}}$:

\noindent{\bf Conjecture 1}: A hard-core nonnegative tempered distribution $g_2({\bf r})$ is a pair correlation function
 of a translationally invariant disordered sphere packing in $\mathbb{R}^d$ at number density $\rho$ 
 for sufficiently large $d$  if and only if
 $S({\bf k})\ge 0$. The maximum achievable density
 is the terminal density $\phi_*$.

In other words, $g_2(r)$ that meets the conditions (i) - (iii),
at or above a critical dimension $d_c$,  packings exist with such a $g_2$. 
A {\it disordered packing} in $\mathbb{R}^d$ is defined in Ref. 9 to be
one in which the pair correlation function $g_2({\bf r})$ decays to its
long-range value of unity faster than $|{\bf r}|^{-d-\epsilon}$ for some
$\epsilon >0$."
Employing the aforementioned optimization procedure with a certain
test function $g_2$ and Conjecture 1, we obtain in what follows
conjectural lower bounds that yield the long-sought
asymptotic exponential improvement on Minkowski's bound.
An important feature of any dense packing is that the particles
form contacts with one another. 
 Experience with disordered jammed packings in low dimensions reveals that the  contact or kissing
number as well as the density can be substantially increased if there
is there is a low probability of finding noncontacting particles from a typical 
particle at radial distances just larger than the nearest-neighbor distance.
It is desired to idealize this small-distance negative correlation (relative to the
uncorrelated value of unity)
in such a way that it is amenable to exact asymptotic analysis.
Accordingly,  a test radial distribution function was considered in Ref. 8
in which there is a gap between the location of a unit step function and the
delta function at finite $d$, i.e.,
\begin{equation}
g_2(r)=\Theta(r-\sigma)+\frac{Z}{s_1(1)\rho}\delta(r-1),
\label{step3}
\end{equation}
where $s(r)$ is the surface area of a  $d$-dimensional sphere of radius $r$ 
and $Z$ is a parameter, which is the average contact or kissing number,
and unity is the sphere diameter. 
The expression contains two adjustable parameters, $\sigma \ge 1$ and $Z$,
which must obviously be constrained to
be nonnegative.

Before reporting the main results of this optimization, it is instructive
to examine the test function (\ref{step3}) for two special cases: (1)
one in which $\sigma=1$ and $Z=0$ and (2) the other in which $\sigma=1$
and $Z>0$ (which were first considered in Ref. 17). In the first special instance,
there are no parameters to be optimized here, and the terminal density $\phi_*$ is given by
$\phi_*=\frac{1}{2^d}$. It is simple to show that the Yamada condition ((\ref{yamada}) is satisfied in any dimension
for $0 \le \phi \le 2^{-d}$.
We already established in the previous section that there exist sphere packings that 
asymptotically have radial distribution functions given by the simple
unit step function for $\phi \le 2^{-d}$. Nonetheless,
invoking Conjecture 1 and the obtained terminal density,
implies the asymptotic lower bound on the maximal density is given by
\begin{equation}
\phi_{\mbox{\scriptsize max}} \ge \frac{1}{2^d},
\end{equation}
which provides an alternate derivation of the elementary bound (\ref{sat}).
Using numerical simulations with a finite but large
number of spheres on the torus, we have been able to construct particle configurations
in which the radial distribution function 
is given by the test function (\ref{step3}) with $\sigma=1$ and $Z=0$ in one, two and three dimensions for densities
up to the terminal density \cite{Cr03,Ou05}. The existence of such
a discrete approximation to this test $g_2$ is suggestive that the standard nonnegativity
conditions may be sufficient to establish existence in this case 
for densities up to $\phi_*$. In the second special case ($\sigma=1$
and $Z>0$) and under the constraint that the minimum of $S(k)$ occurs at $k=0$, then we have the exact results
$\phi_*=\frac{d+2}{2^{d+1}}$ and $Z_*=\frac{d}{2}$, where $Z_*$ is the optimized average kissing number.
The Yamada condition ((\ref{yamada}) is violated here only for $d=1$ 
and becomes less restrictive as the dimension increases from $d=2$.
Interestingly, we have also shown via numerical simulations
that there exist sphere packings possessing radial distribution functions 
given by this test function in two and three dimensions for densities
up to the terminal density \cite{Ou05}. This is suggestive that the Conjecture 1
for this test function may in fact be stronger than is required.
In the high-dimensional limit, invoking  Conjecture 1 and the 
obtained terminal density, yields the conjectural lower bound
\begin{equation}
\phi_{\mbox{\scriptsize max}} \ge \frac{d+2}{2^{d+1}}.
\label{lower-2}
\end{equation}
This lower bound provides the same type of linear improvement over Minkowski's lower
bound as does Ball's rigorous lower bound \cite{Ball92} obtained using
a completely different approach.

Now let us consider the problem when both $\sigma$ and $Z$ in (\ref{step3})
must be optimized. The presence of a gap between the unit step function and delta function will indeed lead
asymptotically to substantially higher terminal densities.
For sufficiently small $d$ ($d \le 200$), the  optimization procedure is carried out numerically \cite{To05b}.
The Yamada condition (\ref{yamada}) is 
violated only for $d=1$ for the test function
(\ref{step3}) for the terminal density $\phi_*$ and associated optimized 
parameters $\sigma_*$ and $Z_*=(2\sigma_* \phi_*)^d -1$. One can
again verify directly that the Yamada condition becomes less restrictive as the
dimension increases from $d=2$. However, although the test function
(\ref{step3}) for $d=2$ with optimized parameters $\phi_*=0.74803$,
$\sigma_*=1.2946$ and $Z_*=4.0148$ satisfies the Yamada condition,
it cannot correspond to a sphere packing because it violates
local geometric constraints specified by $\sigma_*$ and $Z_*$ \cite{To05b}.
To our knowledge, this is the first example of a test radial
distribution function that satisfies the two standard non-negativity
conditions (i) and (iii) and the Yamada condition (\ref{yamada}), but cannot
correspond to a point process.  Thus, there is at least one previously
unarticulated necessary condition that has been violated in the low
dimension $d=2$.   As is the case with the Yamada condition (\ref{yamada}), 
this additional necessary condition appears to lose
relevance in low dimensions because 
 we have shown that there is no analogous local geometric constraint
violation for $d \ge 3$. For $d \le 56$, the terminal density lies below the density of the densest
known packing (a Bravais lattice) \cite{Co93}. However,
for $d >56$,  $\phi_*$ can be larger than the density of the densest
known arrangements, which are ordered. 
Our numerical results for $d$ between 3 and 200, reveal exponential improvement of the terminal density
$\phi_*$ over the one for the gapless case, where $\phi_*=(d+2)/2^{d+1}$.

For large $d$, an exact (but nontrivial) asymptotic analysis can be performed \cite{To05b},
yielding the optimal terminal density. This result in conjunction
with Conjecture 1 yields the conjectural asymptotic lower bound
\begin{equation}
\phi_* \sim
\frac{3.276100896d^{1/6}}{2^{[3-\log_2(e)]d/2}}=\frac{3.276100896
~d^{1/6}}{2^{0.7786524795\ldots \,d}},
\label{den}
\end{equation} 
This putatively provides the long-sought exponential improvement on Minkowski's
lower bound.  We call this a conjectural lower bound because it relies on Conjecture 1 being true,
which a number of results support.
First, the decorrelation principle states that unconstrained
correlations in disordered sphere packings vanish asymptotically in high dimensions
and that the $g_n$ for any $n \ge 3$ can be inferred entirely from a knowledge
of $\rho$ and $g_2$.  Second, the necessary 
Yamada condition appears to only have relevance in very low dimensions.
Third, we have demonstrated that other new necessary conditions
also seem to be germane only in very low dimensions. 
Fourth, we recover the form of known rigorous bounds [cf. (21) and (22)]
in special cases of the test radial distribution function (20) when we
invoke Conjecture 1. Finally, in these two instances,
configurations of disordered sphere packings on the torus 
have been numerically constructed with such $g_2$ in low dimensions for densities up to the terminal density \cite{Cr03,Ou05}. 

A byproduct of the  bound (\ref{den}) is the 
conjectural asymptotic lower bound on the maximal kissing number \cite{To05b}
\begin{equation}
Z_{\mbox{\scriptsize max}} \ge Z_* \sim
40.24850787 \,d^{1/6}\, 2^{[\log_2(e)-1]d/2}=
40.24850787 \,d^{1/6} \, 2^{0.2213475205\ldots\, d},
\end{equation}
This result is superior to the best known 
asymptotic lower bound on the maximal kissing number of $2^{0.2075\ldots d}$
\cite{Wy65}.

The work described above suggests that the densest packings in sufficiently
high dimensions may be disordered rather than periodic, implying
the existence of disordered classical ground states for some continuous potentials.
In fact, there is no fundamental reason why disordered 
ground states are prohibited in low dimensions \cite{Ru82}. A case in point
are the ``pinwheel" tilings of the plane, which possess both statistical translational
and rotational invariance \cite{Ra94}.

This work was supported by the National Science Foundation
under Grant No. DMS-0312067.


\begin{thebibliography}{20}
\bibitem{footnote3}Indeed, we will show in a future
work that the standard RSA packing ($\kappa=0$)
has a substantially higher density than $1/2^d$ in high dimensions.
\bibitem{footnote4} In such dimensions, denser Bravais lattice packings
exist, but it is conjectured that in sufficiently high dimensions
all Bravais lattices become unsaturated (see Ref. \cite{Co93})
and therefore inefficient packings.
\end{thebibliography}
\end{document}